\documentclass[useAMS,usenatbib]{mn2e}
\usepackage{graphicx}
\usepackage[latin1]{inputenc}
\usepackage{subfigure}
\usepackage{indentfirst}
\usepackage{multicol}
\usepackage{Times} 
\usepackage{url}
\usepackage{graphicx}
\usepackage{multicol}
\usepackage{epstopdf}

\title[The journey of Typhon-Echidna as a binary system through the planetary region]
{The journey of Typhon-Echidna as a binary system through the planetary region}
\author[R. A. N. Araujo, M. A. Galiazzo, O. C. Winter, R. Sfair]
{R. A. N.  Araujo$^{1}$\thanks{E-mail: ran.araujo@gmail.com}, 
M. A. Galiazzo$^{2}$\thanks{E-mail: mattia.galiazzo@univie.ac.at},  
O. C. Winter$^1$\thanks{E-mail: ocwinter@gmail.com}, 
R. Sfair$^1$\thanks{E-mail: rsfair@feg.unesp.br}\\ 
$^{1}$UNESP - S\~ao Paulo State University, Grupo de Din\^amica Orbital e Planetologia, CEP 12516-410, Guaratinguet\'a, SP, Brazil \\
$^{2}$Institute of Astrophysics, University of Vienna, A-1180 Vienna, Austria
}

\begin{document}
\date{}
\pagerange{\pageref{firstpage}--\pageref{lastpage}} \pubyear{2017}
\maketitle
\label{firstpage}

\begin{abstract}
 Among the current population of the 81 known trans-Neptunian binaries (TNBs), only two are in orbits that 
cross the orbit of Neptune. These are (42355) Typhon-Echidna and (65489) Ceto-Phorcys.  In the present work, we 
focused our analyses on the temporal evolution of the Typhon-Echidna binary system through the outer and inner planetary 
systems. Using numerical integrations of the N-body gravitational problem, we explored the orbital evolutions of 500 
clones of Typhon, recording the close encounters of those clones with planets. We then analysed the effects of those 
encounters on the binary system. It was found that only $\approx22\%$ of the encounters with the giant planets were 
strong enough to disrupt the binary. 
This binary system has an $\approx3.6\%$ probability of reaching the terrestrial planetary region over a time scale 
of approximately $5.4$ Myr. Close encounters of Typhon-Echidna with Earth and Venus were also registered, but the 
probabilities of such events occurring are low $(\approx0.4\%)$. The orbital evolution of the system in the past was also investigated. 
It was found that in the last $100$ Myr, Typhon might have spent most of its time as a TNB crossing the orbit of Neptune. 
Therefore, our study of the Typhon-Echidna orbital evolution illustrates the possibility of large cometary bodies (radii of $76$ km for Typhon and $42$ km for Echidna) coming from a 
remote region of the outer Solar System and that might enter the terrestrial planetary region preserving its binarity throughout the journey.
\end{abstract}


\section{Introduction}

\begin{table*}
 \begin{center}
 \begin{minipage}{9cm}
\caption{Orbital and Physical parameters of the Typhon-Echidna binary system.}
\end{minipage}
  \begin{tabular}{c c c c c c c}
  \hline
   Body		 &Orbits	&$a_0$			&$e_0$  	&$i_0$  		&Mass$^{3}$ (kg)	&Radius$^{2}$ (km)  \\
		 &		&			&		&			&			&   \\
   \hline
   Typhon  $^{1}$&Sun		&38.19520 au		&0.54115	&2.42776$^{\circ}$	&$8.1 \times 10^{17}$	&76	\\
   Echidna $^{2}$&Typhon	&1628 km	&0.526		&37.9$^{\circ}$		&$1.4 \times 10^{17}$	&42 	\\
   \hline
     \multicolumn{7}{l}{$^{(1)}$Orbital elements of Typhon obtained through HORIZONS Web-Interface at 2453724.5 JDTBT}\\
  \multicolumn{7}{l}{$^{(2)}$ Obtained from \cite{grundy}}\\
  \multicolumn{7}{l}{$^{(3)}$ Calculated assuming spherical bodies and a density of $0.44g/cm^3$. This density was estimated by}\\
   \multicolumn{7}{l}{\cite{grundy}, considering that the total mass of the system was $9.49\times 10^{17}$ kg as well as the radii of the bodies.}\\
 \label{tab_data}
  \end{tabular}
 \end{center}
\end{table*}

\begin{table}
\begin{center}
\caption{Tidal disruption radius $(r_{td})$ and the registered close encounters of Typhon and its clones along the numerical integration ($<200$ Myr).}
\begin{tabular}{c c c c c}
\hline
Planet 	        &1 $r_{td}$ 			&1 $r_{td}$	  	&Significant 				&Extreme 		\\
		&$(\times 10^5 km)$		&Planet's radius	&Encounters				&Encounters	\\
		& 				&			&$d \leq 10\hspace{0.1cm}r_{td}$	&$d \leq 3\hspace{0.1cm}r_{td}$   \\
\hline
Venus		& $4.04$			&$67$	 		&36					&5				 \\
Earth 		& $4.34$			&$68$			&46					&3			\\
Mars  		& $2.09$ 			&$62$			&4					&0	\\
Jupiter 	& $29.6$			&$41$			&1889					&317	\\
Saturn 		& $19.73$			&$33$			&640					&101	\\
Uranus 		& $10.47$			&$40$			&912					&120	\\
Neptune 	& $11.21$			&$44$			&589					&81	\\
\hline
\multicolumn{5}{l}{d - minimum distance of the close encounter}\\
\end{tabular}
\label{tab_enc}
\end{center}
\end{table}

There is no consensus on the definition of Centaurs (as discussed in \citealt{araujo2016}), but it is commonly accepted that they have a perihelion between the orbits of the giant planets. According to Johnston, W.R.\footnote{http://www.johnstonsarchive.net/astro/tnos.html}, Centaurs are bodies that cross the orbit of Neptune (with a perihelion distance of less than $30$ au), while TNOs are bodies with semimajor axes greater than the semimajor axis of Neptune ($a>30$ au). Based on these definitions, we consider Centaurs as bodies in orbits with perihelion distances of less than $30$ au and with $a\leqslant30$ au, and the TNOs as bodies in orbits with perihelion distances greater than $30$ au and with $a>30$ au. The intersection between TNOs and Centaurs, i.e., bodies with $a>30$ au and with perihelion distances less than $30$ au, are referred to herein as TNO-Centaurs. The resonant bodies with $a>30$ au and $q<30$ au are not included in these definitions since they generally present stable orbits and do not currently suffer close encounters with Neptune or another planet.
 
There are studies showing that trans-Neptunian objects (TNOs) can evolve to become 
Centaurs \citep{Lev1997,tiscareno,lykawka,disisto2009,brasser2012}. TNOs are also among the sources of near-Earth 
objects (NEOs) \citep{Mor1997,Lev1997,tiscareno,disisto2007,stell2014}. According to \cite{morbidelli2002}, 
$6\pm4~\%$ of NEOs come from the trans-Neptunian region. \citet{Gal2016} found that TNOs, at least those whose orbits have perihelia 
smaller than $34$ au, can become Centaurs and may evolves within the inner solar system. According to \cite{napier2015}, 
approximately one in ten Centaurs in (2060) Chiron-like orbits enter the Earth-crossing region. 

The evolutions of the trajectories of small bodies suffering close encounters with the giant planets generate orbital instabilities, which characterize chaotic motion \citep{araujo2016, Lev1997, Mor1997, Hor2004a, Hor2004b, Gal2016}. Recently, a Centaur, (10199) Chariklo, was found to have a well-defined ring system \citep{braga}. Despite the numerous close encounters with giant planets, the ring systems have proven to be highly stable \citep{araujo2016} along Chariklo's lifetime as a Centaur, which is approximately 10 Myr \citep{Hor2004a}.

According to \cite{fraser}, planetesimals formed near the Kuiper Belt are expected to form as binaries. Trans-Neptunian binaries (TNBs) are estimated to be $\sim20$\% of all TNOs \citep{Nol2008}. Today, 81 TNBs are known\footnote{http://www.johnstonsarchive.net/astro/astmoontable.html}, with cases of multiple systems, e.g., Pluto, Haumea and 1999 TC36. Knowing that TNOs evolve to become Centaurs or even NEOs, the question that arises and that we aim the answer is how would the evolution of a binary TNO proceed when entering the outer or even the inner planetary region?

Currently there are only two known binary TNO-Centaurs: (42355) Typhon-Echidna and (65489) Ceto-Phorcys. The current orbits of the other 79 TNBs behave as typical TNOs, i.e., with no crossing of planetary orbits. Typhon was the first binary TNO-Centaur discovered \citep{Nol2006a}; its secondary is named Echidna. Typhon has heliocentric orbital elements $a = 37.9$ au, $e = 0.538$, and $i = 2.43^\circ$ \citep{grundy} and a perihelion distance of $q=17.5$ au. Ceto's binarity was discovered in the same year as Typhon's \citep{Nol2006b}; its secondary is named Phorcys. Ceto has heliocentric orbital elements $a = 102$ au, $e = 0.82$, and $i = 22^\circ$ \citep{grundy_ceto} and a perihelion distance of $q=18.4$ au. Note that Ceto has a small portion of its orbit inside the orbit of Neptune, since its semi-major axis is more than three times larger than that of Neptune. Therefore, due to the size of its semi-major axis and due to its high orbital inclination, Ceto is expected to experience a lower frequency of close encounters with the giant planets \citep{araujo2017} than is Typhon.

The Typhon-Echidna system was then chosen since, in the current work, we are interested in studying a binary TNO-Centaur that evolves mostly within the planetary region. First, the heliocentric orbital evolution of Typhon was explored. Through numerical integrations of the equations of the N-body gravitational problem, close encounters of clones of Typhon with the planets were recorded in the model integrations. We then analysed the effects of those registered encounters on the binary system.

In the next section, the study of the orbital evolution of Typhon is presented, followed by the exploration of the effects of close encounters on the binary Typhon-Echidna. In Section \ref{sec_terrestre}, the possibility of Typhon entering the terrestrial planetary region as a binary system is discussed. A brief study of the past evolution of Typhon-Echidna is shown in Section \ref{sec_past}. Our final comments are presented in the last section.

\begin{figure}
\begin{center}
\includegraphics[scale=0.52]{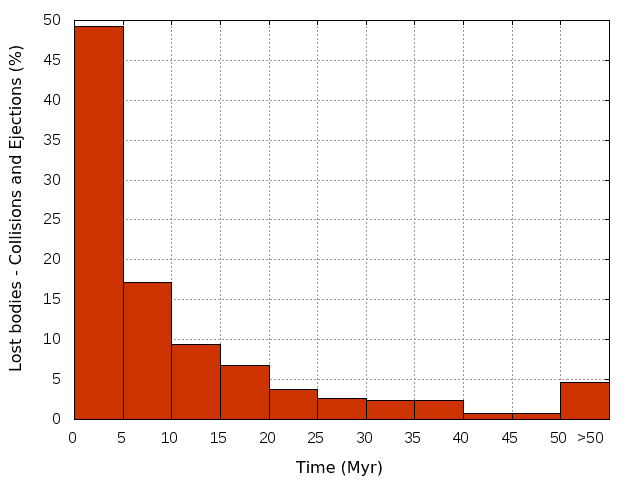}
\caption{Loss of bodies as a function of time for Typhon and its clones.}
\label{fig_lifetime}
\end{center}
\end{figure}

\begin{figure*}
\begin{center}
\subfigure[]{\includegraphics[scale=0.65]{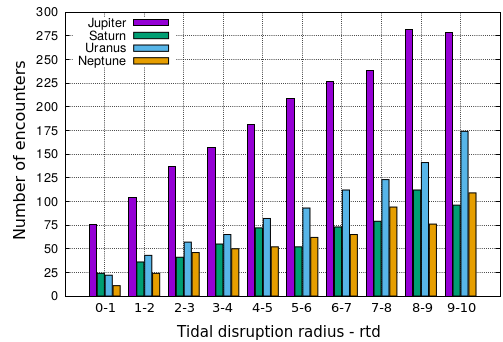}}
\subfigure[]{\includegraphics[scale=0.65]{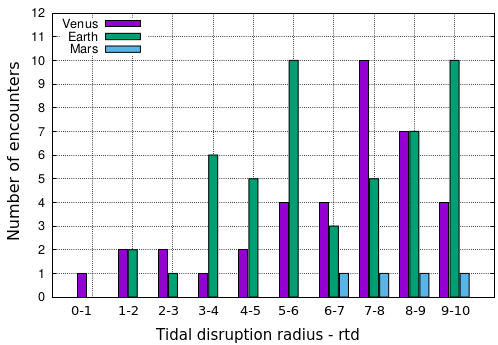}}
\caption{Number of close encounter inside each range of the tidal disruption radius of each planet. a) Giant planets. b) Terrestrial planets.
The encounters were computed along the numerical integration ($<200$ Myr).}
\label{fig_rtd}
\end{center}
\end{figure*}

\section{Numerical Simulations}
\label{sec_num_simula}
The numerical simulations were performed in two steps. First, we simulated the orbits of a set of clones of Typhon. 
In this step, we aimed to analyse the orbital evolution of the clones as they evolve from their initial orbit. 
We also registered the close encounters of the clones with the giant or with the terrestrial planets. 
The next step consisted of simulating a selection of the most significant close encounters 
(as defined in Section \ref{sec_orb_evol}) considering that, at this time, a binary system is experiencing the encounter. 
We then analysed the effects of these encounters in terms of their disruptions of the binary.

\subsection{Orbital Evolution}
\label{sec_orb_evol}

The orbital and physical data of the Typhon-Echidna binary are presented in Table \ref{tab_data}.
Assuming those parameters, we simulated a set of $500$ clones of Typhon. The clones are massless bodies with small deviations in their orbits. They were generated as in \cite{Hor2004a}, i.e., by assuming an interval of variation of $a=a_0\pm 0.005$ au for the semi-major axis, $e=e_0\pm 0.003$ for the eccentricity and  $i=i_0\pm 0.01^{\circ}$ for the orbital inclination, following a Lorentzian distribution centred at the initial osculating elements ($a_0$, $e_0$ and $i_0$). The angular elements $\omega=158.98^{\circ}$, $\Omega=351.99^{\circ}$ and $f=358.08^{\circ}$ (argument of perihelion, longitude of the ascending node and true-anomaly, respectively) are the same for all the clones and were obtained for Typhon through the HORIZONS Web-Interface at 2453724.5 JDTBT (in Julian Date, Barycentric Dynamical Time).

A simulation of a system comprising the Sun, the $500$ small bodies and all the planets of the Solar System, excluding Mercury and including the dwarf planet Pluto, was considered. The orbits of the planets were also taken from the HORIZONS Web-Interface for the same Epoch.

The numerical integration was performed using the adaptive time-step hybrid sympletic/Bulirsch-Stoer algorithm from \textsc{Mercury} \citep{chambers} for a time span of $200$ Myr, time step of 2 days and with the output interval of the data being 2000 years. Throughout the integration, the clones did not interact with each other, but they could collide with the massive bodies or be ejected. 
A clone was considered ejected if its orbital radius reached $110$ au. 
This value was adopted taking into account that if a clone reach the distance of $110$ au and its orbital eccentricity is lower than 
$0.6$, then it is already a TNO ($q>30$ au and $a>30$ au), while clones reaching $110$ au with higher eccentricity $(e>0.6)$ spend only a small fraction of their orbital period within the planetary region.
The collisions were defined by the physical radii of the planets. The collisions with the Sun were defined by the distance of $0.009$ au. This value corresponds to approximately two times the radius of the Sun.
Occasionally there could be some clones approaching the Sun at perihelion distances smaller than $\approx 0.1$ au. The temporal evolution of those clones may gradually suffer from orbital inaccuracies when using the hybrid algorithm with the assumed time step. However, as it will be seen in Section \ref{sec_terrestre}, only few clones may have come so close to the Sun and thus, such presumable inaccuracy on their orbital evolution do not compromise our statistical analyses.

As a result of this step, we found that $100\%$ of the clones were lost along the timespan of the integration. Only $4$ clones were lost via collisions, including $2$ collisions with Uranus, $1$ collision with Jupiter and $1$ collision with the Sun. The other clones were ejected. The clone that survived the longest was ejected after $163$ Myr. The histogram in Fig. \ref{fig_lifetime} shows the loss of clones as a function of time. Approximately $80\%$ of the clones do not survive more than $20$ Myr. In fact, we found that $50\%$ of the clones survived  only slightly longer than $5$ Myr. The calculated median gives the estimated lifetime of Typhon as $5.2$ Myr. This value is shorter than the mean value of approximately $10$ Myr found by \cite{tiscareno} for the lifetime of Centaurs, which was determined considering a sample of 53 objects with perihelion distances within the orbit of Neptune, including Typhon. However, the authors emphasize that they found a wide variety of lifetimes, ranging from less than $1$ Myr to more than $100$ Myr.

All close encounters experienced by the clones with any of the planets within the 
distance of $10$ $r_{td}$ and along the numerical integration ($<200$ Myr) were registered. The tidal disruption radius $(r_{td})$ provides an approximate distance for which a given binary is expected to be disrupted due to the effects of a nearby gravitational encounter with a more massive body \citep{philpott}. The expression for this is given as follows:

\begin{equation}
 r_{td}  \approx a_B\left(\frac{3M_p}{M_1+M_2}\right)^{1/3}
 \label{eq_rtd}
\end{equation}
where $M_p$ is the mass of the encountered planet, $M_1+M_2$ is the total mass of the binary and $a_B$ is the separation of the binary.

In Table \ref{tab_enc}, we present the $r_{td}$ calculated for the Typhon-Echidna system and the number of registered close encounters within $10~r_{td}$ and $3~r_{td}$ observed by this system when each of the planets is considered. Hereafter, the close encounters within $10~r_{td}$ are called significant encounters. The closest encounters experienced within $3~r_{td}$ are the so-called extreme encounters. Fig. \ref{fig_rtd} shows the distribution of the registered close encounters by planets and for a range of values of $r_{td}$.

From Fig. \ref{fig_rtd} and Table \ref{tab_enc} we see that, among the giant planets, the most encountered planet is Jupiter, followed by Uranus, Saturn and Neptune. We see that there were also registered significant and extreme encounters of Typhon with the terrestrial planets. This fact will be further discussed in Section \ref{sec_terrestre}. At this point, we seek to analyse how the encounters with the giant planets perturb the binary system. The discussion and results of this analysis are presented in the following section.

\subsection{Binary Evolution}
\label{sec_bin_evol}

\begin{figure}
\begin{center}
\subfigure{\includegraphics[scale=0.51]{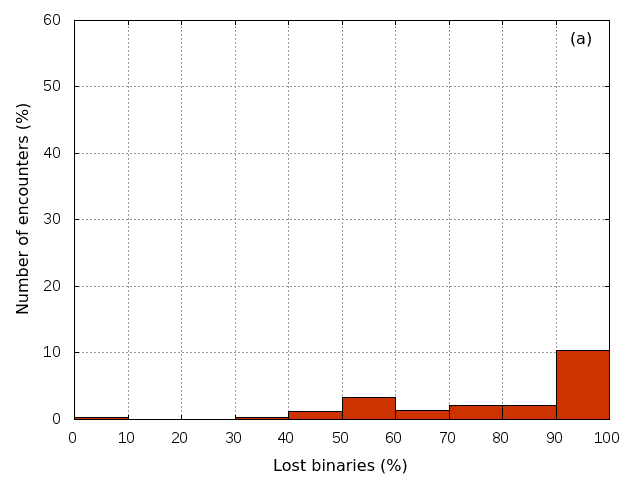}}
\subfigure{\includegraphics[scale=0.51]{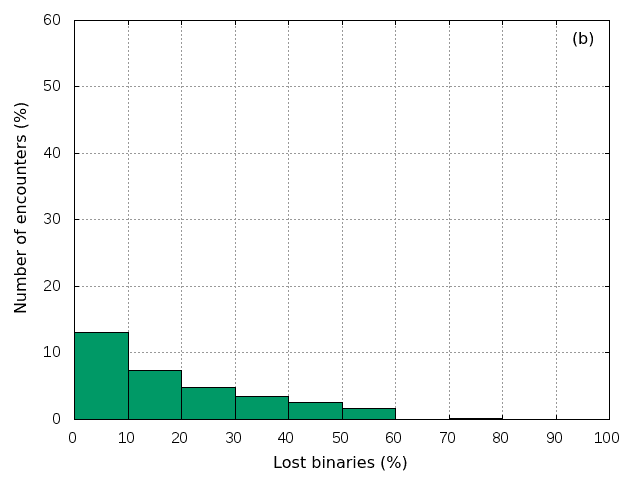}}
\subfigure{\includegraphics[scale=0.51]{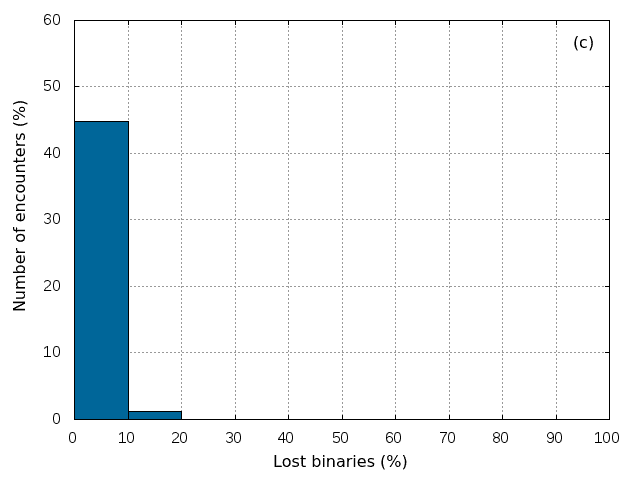}}
\caption{Percentage of binaries lost (relative to the total of $12960$ Typhon-Echidna-like binary systems), 
as a function of the percentage of 
the extreme encounters (relative to the total of 619 extreme encounters with the giant planets) performed 
within: a) $1~r_{td}$,  b) $2~r_{td}$ and  c) $3~r_{td}$.}
\label{fig_binaries_lost}
\end{center}
\end{figure}

\begin{figure}
\begin{center}
\subfigure{\includegraphics[scale=0.5]{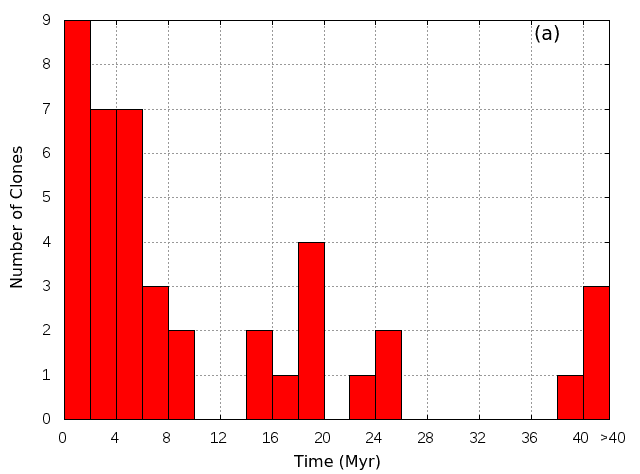}}
\subfigure{\includegraphics[scale=0.5]{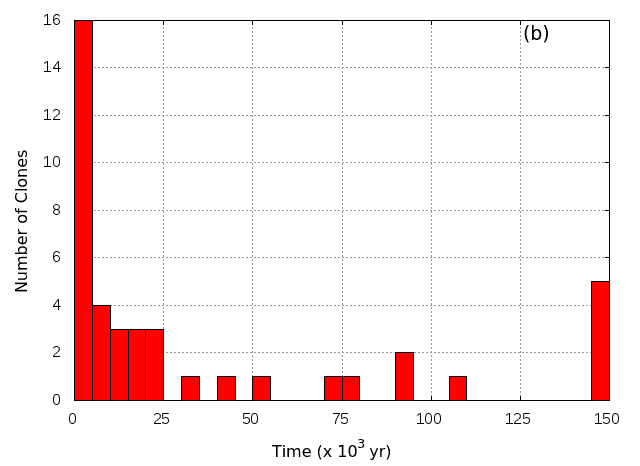}}
\subfigure{\includegraphics[scale=0.5]{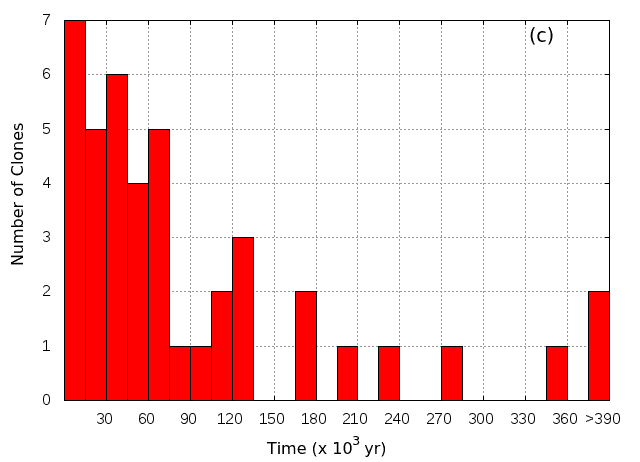}}
\caption{a) Time needed for the 42 clones to reach the terrestrial region from the current orbit of Typhon. 
 b) Time spent between the first and the last times that the distance between the Sun and each one of the clones was within the limit of $2$ au.
 c) Survival time of the clones after leaving the terrestrial region, i.e., crossing orbits.}
\label{fig_clones_terrest}
\end{center}
\end{figure}

\begin{figure*}
\begin{center}
\subfigure[]{\includegraphics[scale=0.4]{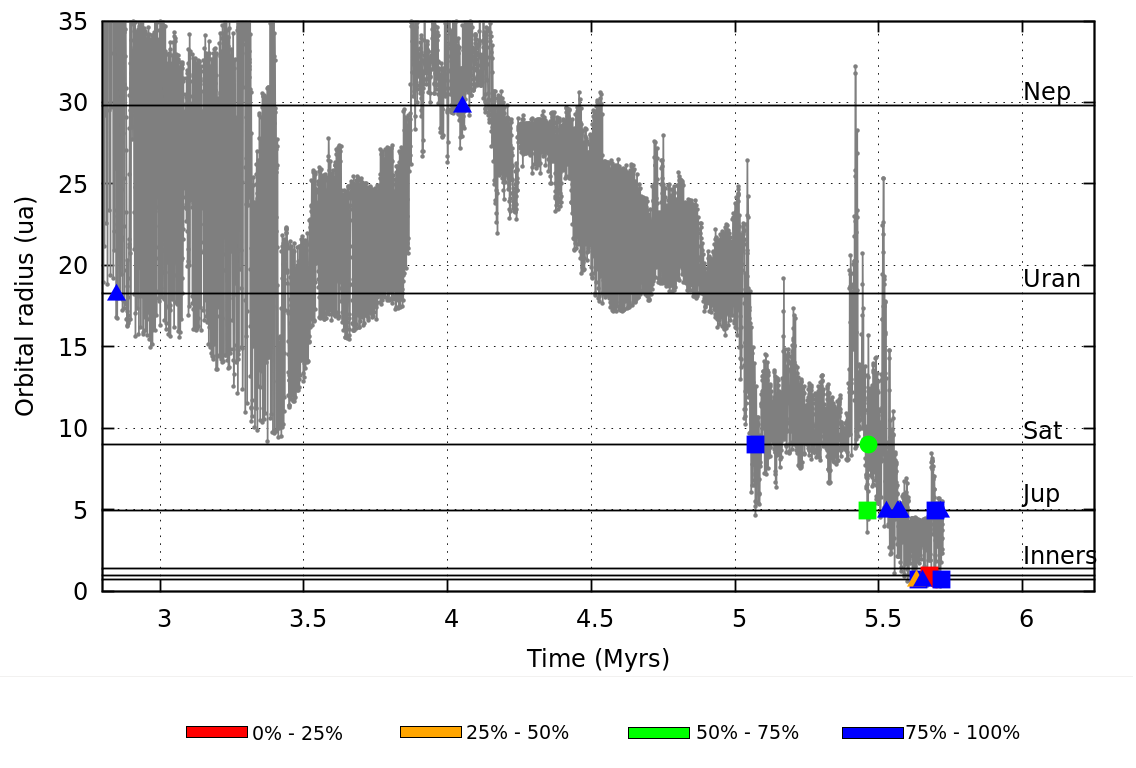}}
\subfigure[]{\includegraphics[scale=0.4]{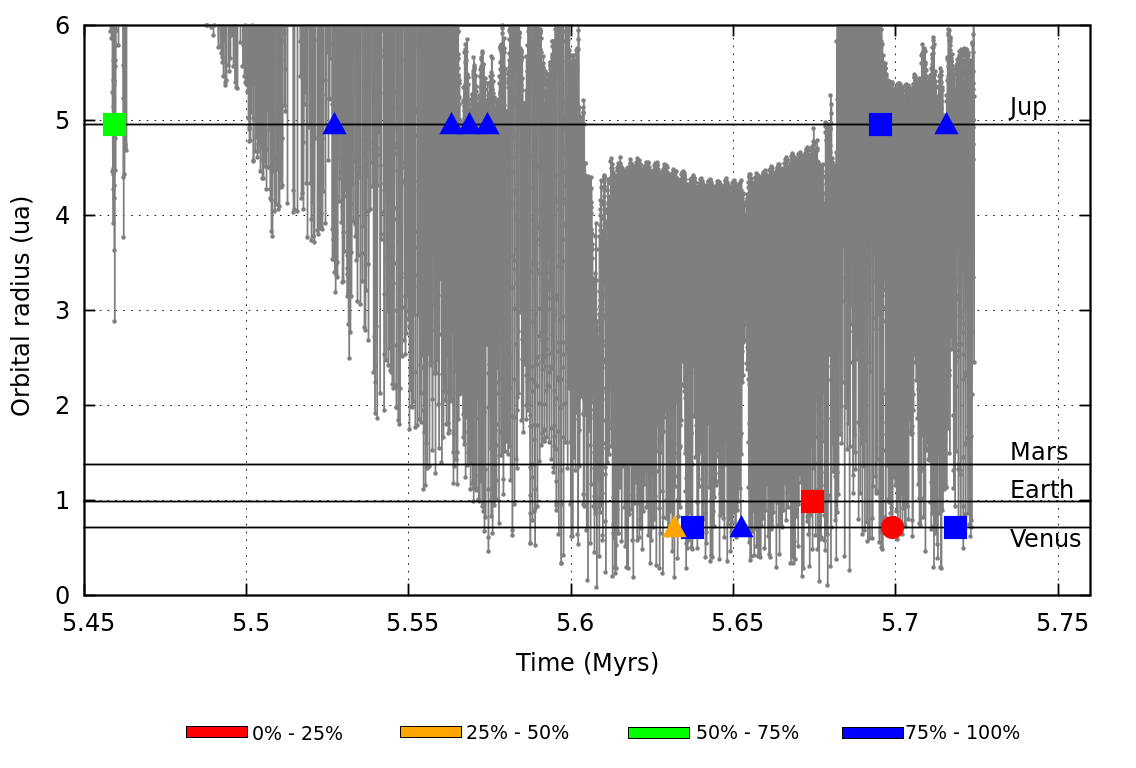}}
\caption{Orbital evolution and the extreme encounters of a Typhon-Echidna system that reached the region of the terrestrial planets. 
The colour-coded label indicates the percentage of binary systems lost due to close encounters relative to the total set of $12960$ binaries.
Triangles show encounters at $2 < r_{td} \leq 3$. Squares show encounters within $1< r_{td} \leq 2$, and circles show encounters 
with $r_{td} \leq 1$.
The grey line indicates the orbital radius of the system on its heliocentric orbit.
a) For the entire interval of time during which the encounters within $3\,r_{td}$ were registered. 
b) Zoom showing mainly the extreme encounters with the terrestrial planets.}
\label{fig_evolution}
\end{center}
\end{figure*}

\begin{figure*}
\begin{center}
\includegraphics[scale=0.4]{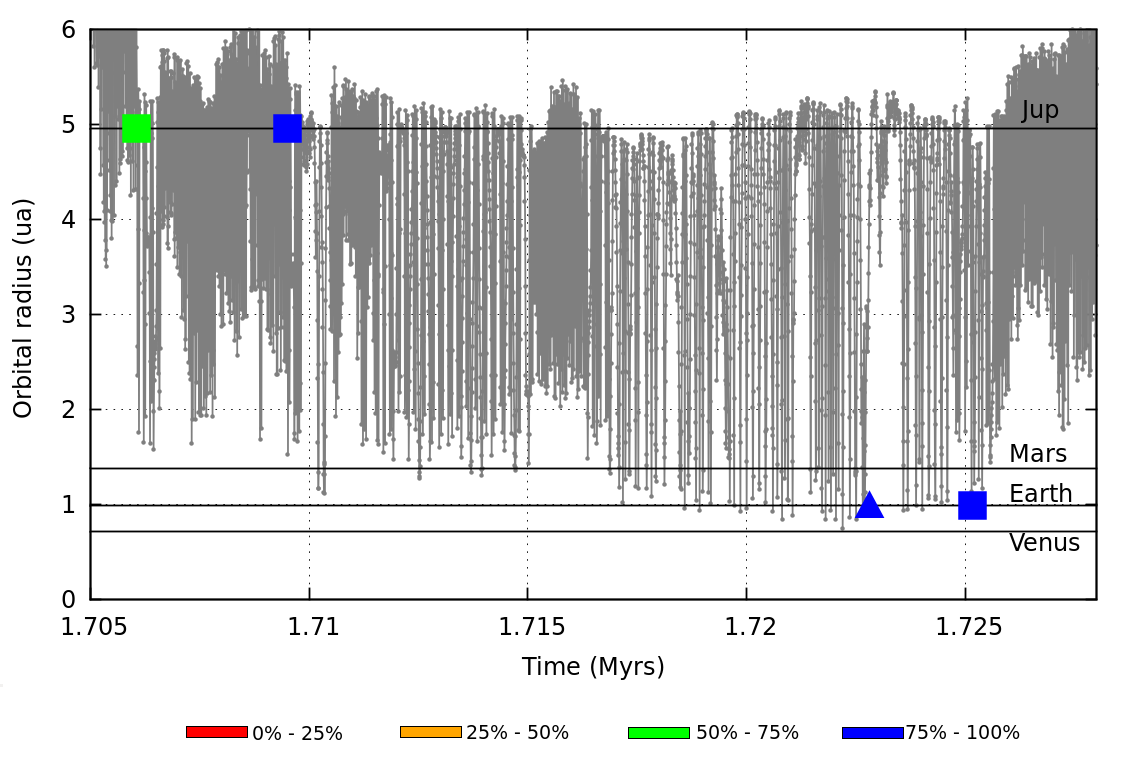}
\caption{Orbital evolution and the extreme encounters of a Typhon-Echidna system that reached the region of the terrestrial planets 
for the entire interval of time in which the encounters within $3\,r_{td}$ were registered.
The colour-coded label indicates the percentage of binary system lost due to the close encounters relative to the total set of $12960$ binaries.
Triangles show encounters at $2< r_{td} \leq 3$. Squares show encounters within $1< r_{td} \leq 2$, and circles show encounters 
with $r_{td} \leq 1$.
The grey line indicates the orbital radius of the system on its heliocentric orbit.}
\label{fig_evolution2}
\end{center}
\end{figure*}

\cite{araujo2016} studied the problem of the rings of Chariklo perturbed by close gravitational encounters with the giant planets. They found that the encounters capable of producing any significant effects on the rings occurred at distances smaller than $2~r_{td}$. Based on their results, we selected those encounters that occurred within $3\,r_{td}$ among all the significant close encounters registered throughout the numerical integrations, i.e., we selected the extreme close encounters. The number of extreme encounters of the clones with the planets is presented in Table \ref{tab_enc}, column $4$, from which we see that we have a total $627$ extreme close encounters to analyse. Each one of these $627$ extreme encounters was individually simulated considering a system comprising the Sun, the planet, the clone of Typhon involved in the encounter and the secondary component Echidna.

The initial conditions of the planet and clone were obtained from the previous numerical integrations (Section \ref{sec_orb_evol}). In these simulations, every time that a significant close encounter was registered, we recorded the components of the heliocentric orbital positions and of the orbital velocities of the clone and planet involved in the encounter. These data were recorded at the moment prior to the nearest crossing of the encounter at $10~r_{td}$. Thus, here, we analysed the effects on the binary system due to the encounters with the minimum distance of the encounter being within $3~r_{td}$ (extreme encounters). However, all the simulations of the encounters started at relative distances on the edge of $10~r_{td}$ of each planet.

Echidna was distributed around Typhon, which was treated as a point mass, with the following initial conditions: the semi-major axis was $a=1628$ km, eccentricity was $e=0.526$, orbital inclination was $0^{\circ} \leq i \leq 180^{\circ}$ with steps of $45^{\circ}$, longitude of the ascending node was at $\Omega=0^{\circ}$ and $\Omega=90^{\circ}$, argument of perihelion was $0^{\circ} \leq \omega \leq 360^{\circ}$ with steps of $10^{\circ}$ and true anomaly was $0^{\circ} \leq f \leq 360^{\circ}$ with steps of $10^{\circ}$. The orbital inclination of Echidna was varied to simulate different positions of the equator of Typhon at the moment of the close encounter. 
By creating a 3D-cloud of Echidnas around Typhon, we considered a large range of possibilities for the geometry of the encounter (via the position of Typhon-Echidna binary system relative to the encountered planet). The combination of these values resulted in a cloud with $12960$ Typhon-Echidna-like systems performing each of the registered extreme close encounters. This approach allows us to statistically analyse the effects of the close encounters while taking into account different geometries of the binary system during the encounter.

Here, Typhon and Echidna were considered massive bodies such that the binarity of the system was preserved. However, the Echidnas do not interact with each other. The cumulative effect of the extreme encounters on a single binary was also not taken into account. Our study consisted of a statistical analysis based on the number of extreme encounters and how they are expected to affect a Typhon-Echidna like system. 

At this step, numerical integrations were performed using the adaptive time-step Gauss-Radau numerical integrator \citep{everhart} for a time span of $1$ year. Throughout the simulation of the extreme encounters, the Echidnas could collide with Typhon, or the binary could be disrupted. A collision was defined by the physical radius of Typhon and the disruption was defined by the two-body energy of the system comprising Typhon and Echidna. When this energy was initially negative but became positive, we computed a disruption of the Typhon-Echidna system.
 
After the numerical integrations, we analysed, for each extreme close encounter, the percentage loss (via collisions or disruptions) of the binaries relative to the initial total number of $12960$ binaries per encounter. The results of the encounters with the giant planets are presented in Fig. \ref{fig_binaries_lost}. The analysis for the terrestrial planets is presented in the next section. The graphs of this figure show the percentage of binaries lost as a function of the percentage of extreme encounters (relative to the total number of 619 extreme encounters with the giant planets). For a better visualization, these results were presented separately according to the distances of the encounters. Fig. \ref{fig_binaries_lost}a shows the results of the encounters at distances $d\le1~r_{td}$, Fig. \ref{fig_binaries_lost}b shows the encounters at distances $1< d\le2~r_{td}$ and Fig. \ref{fig_binaries_lost}c shows the encounters at distances $2< d\le3~r_{td}$.

We assume that an extreme close encounter is capable of disrupting the Typhon-Echidna binary system if more than $50\%$ of the binaries from the cloud of $12960$ binaries are lost. We refer to these encounters as disruptive encounters. Under this assumption, from Fig. \ref{fig_binaries_lost}, it is confirmed that, the disruptive encounters mainly occurred within $1~r_{td}$.  From the total of 619 extreme close encounters with the giant planets simulated, $133$  (i.e.,$133/619\approx21.5\%$) occurred within $1r_{td}$, and 129 (i.e., $129/133\approx97\%$) were disruptive encounters. We also computed 11 disruptive encounters (i.e., $11/619\approx1.8\%$ of the extreme close encounters) that occurred within the limit of $1< d\le2~r_{td}$.  None of the encounters within $2< d\le3~r_{td}$ were disruptive (\ref{fig_binaries_lost} c). These results led to estimations that, from the total of 619 extreme close encounters of Typhon with any of the giant planets, only 140 (i.e., $140/619\approx23\%$ of the extreme close encounters) were disruptive, i.e., capable of disrupting the Typhon-Echidna system. Thus, our simulations of the binary system of Typhon-Echidna experiencing extreme close encounters with the giant planets showed that most of these encounters are harmless to the system.

Another way to analyse these data is to compute the number of disruptive encounters relative to the number of clones that experienced them. This was done by computing the close encounters of each clone within $1r_{td}$, since these encounters were shown to be almost exclusively responsible for the disruptive encounters, as discussed above. We found that the 133 encounters that occurred within $1r_{td}$ were experienced by $84$ clones. Eighteen clones suffered a total of $67$ multiple encounters within $1r_{td}$, while $66$  clones suffered only one encounter within $1r_{td}$. This number leads us to estimate that approximately $17\%~(\approx84/500)$ of the clones would lose their second component due to close gravitational
encounters with the giant planets. This analysis confirms our previous conclusion. The binary system of Typhon-Echidna is more likely to survive close gravitational encounters with these planets.

However, as previously stated, the small bodies of the outer Solar System called Centaurs or TNOs are known to be able to reach the region of the terrestrial planets, and as presented in Table \ref{tab_enc}, we have registered some significant and extreme encounters of Typhon with Mars, Earth and Venus. Thus, if Typhon-Echidna is more likely to safely pass through the region of influence of the giant planets and if the orbit of this system is such that it can reach the inner Solar System, is it possible to have a binary system as large as this one transiting near us? Note that the sizes of the bodies of Typhon and Echidna are an order of magnitude larger than the largest NEOs. This is one of the questions addressed in the following section.

\section{Typhon-Echidna into the terrestrial planetary region}
\label{sec_terrestre}

\begin{figure}
\begin{center}
\subfigure[]{\includegraphics[scale=0.5]{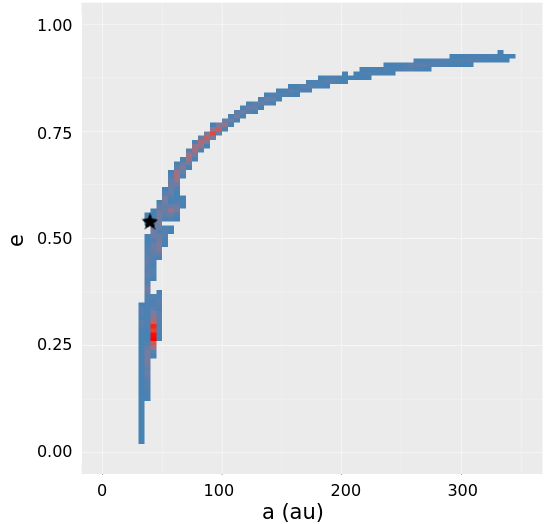}}
\subfigure[]{\includegraphics[scale=0.5]{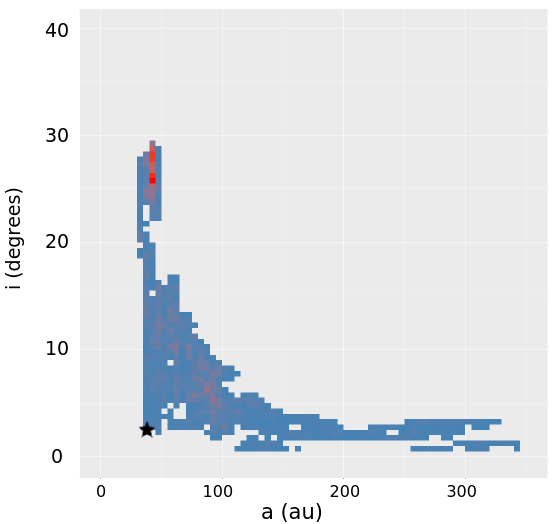}}
\caption{Example of the backward temporal evolution of a clone of Typhon in a) $(a\times e)$ space, b) $(a\times i)$ space. 
The clone in this example started the integration with $a=38.19460$ au, $e=0.54090$ and $i=2.42670^{\circ}$ (black stars) and survived (no ejection or collision) for the 
entire integration time of $100$ Myr to the past.
The colour-coding indicates the time spent by the clone in a given region of the spaces from shorter (light blue) 
to longer (dark red). The graphs show that the clone spent the most time with $a=42.5$ au, $e=0.265$ and $i=27.8^{\circ}$.
} 
\label{fig_example}
\end{center}
\end{figure}

\begin{figure}
\begin{center}
\subfigure[]{\includegraphics[scale=0.45]{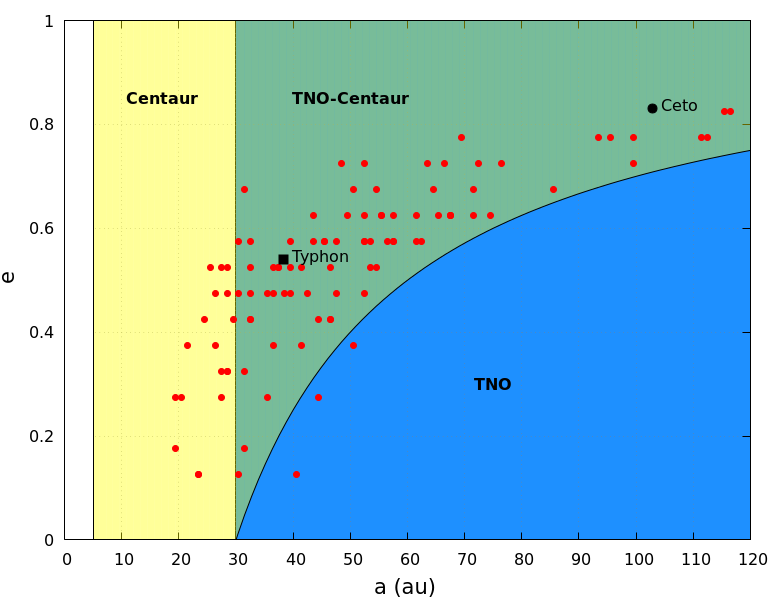}}
\subfigure[]{\includegraphics[scale=0.44]{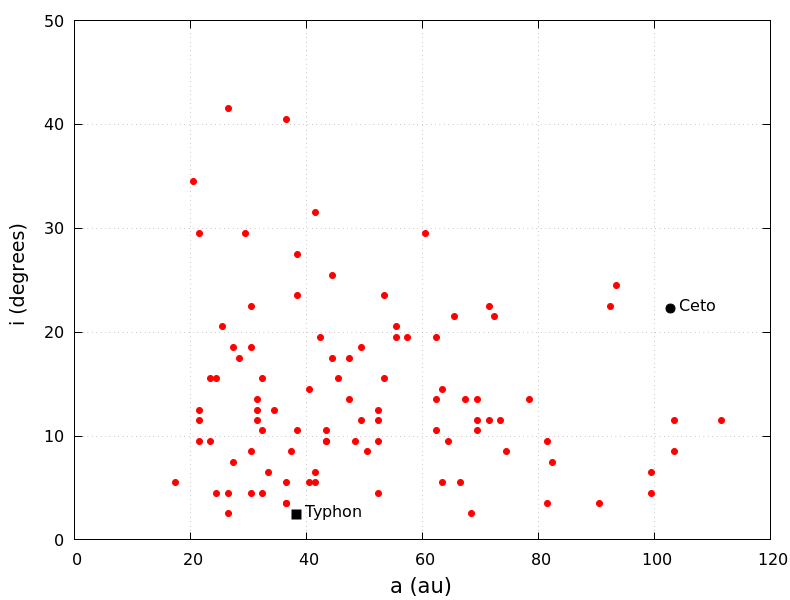}}
\caption{Backward temporal evolution of the semi-major axis $a$, eccentricity $e$ and inclination $i$ for $100$ clones of Typhon.
The locations in a) $(a\times e)$ space and b) $(a\times i)$ space are shown, which is where the clones spent most of the time during their integration.
In the first graph, the Centaur region (yellow region) is defined by the semimajor axis $a<30$ au and $q<30$ au, and 
the TNO region (blue region) is defined by a perihelion distance greater than $30$ au (with no crossing of Neptune's orbit). 
The TNO-Centaur region is the intersection between the Centaurs and the TNOs (green region) with $a>30$ au and a perihelion distance less than $30$ au.
The current orbits of Typhon and Ceto are indicated in the graphs by the black square and by the black circle, respectively.} 
\label{fig_past}
\end{center}
\end{figure}

During the numerical simulations of the $500$ clones of Typhon (Section \ref{sec_orb_evol}), we followed all clones that reached the terrestrial planetary region. We monitored the clones whose orbital radii were smaller than $2$ au by computing the relative distances of the clones to the Sun for each time step throughout the integration.
These are candidates that might experience significant and extreme encounters with the terrestrial planets since, at the heliocentric distance of $2$ au, the clones have already left the inner border of the Main Belt Asteroid \citep {demeo} and are about to become Mars-crossers \citep{michel}.

From the total of $500$ clones, we found that $42$ reached this region, implying a probability of $8.4\%$ of Typhon moving into the region of the terrestrial planets along the numerical integration over $<200$ Myr. Those clones suffered a total of $86$ significant encounters with the planets Mars, Earth and Venus (see Table \ref{tab_enc} and Fig. \ref{fig_rtd}b). 

By examining the dynamical evolutions of those $42$ clones before their entries, we found that $23$ suffered close encounters with a giant planet within $1\,rtd$. Those are the encounters capable of disrupting the binary system, as discussed in Section \ref{sec_bin_evol}. Thus, 19 clones ($\approx45\%$ from the total of 42 clones that reach the terrestrial region) are likely to be accompanied by the secondary component Echidna. Considering these 19 clones relative to the total sample of $500$ clones, we estimate that the Typhon-Echidna system has a $3.8\%$ probability of reaching the terrestrial region as a binary.

We then analysed how long it took for each of the $42$ clones to reach the terrestrial region. The entry time is the time computed since the beginning of the integrations until the clone crosses the limit of $2$ au. The first clone to reach the terrestrial region took $\approx4.27\times10^{5}$ years. The last occurred at $\sim6.6\times10^{7}$ years. The distribution of the entry time for the $42$ clones is presented in 
Fig. \ref{fig_clones_terrest}a. 
The calculated median shows that the clones are more probable to reach the terrestrial region in $\approx 5.4$ Myr. 
On average, this region is achieved by them in $\approx 12.2$ Myr. This value is consistent with the average time for a wider sample of Centaurs and TNOs entering the main asteroid belt, as obtained by \citet{Gal2016}, which is estimated to be $10.0$ Myr for the Centaurs and $16.8$ Myr for the TNOs.
In this work, the authors considered the orbits of known Centaurs and TNOs with $5.5\leq a \leq 80$ au and perihelion distances of less than 40 au. This range includes Typhon.

The histogram in Fig. \ref{fig_clones_terrest}b shows the time spent between the first and the last times that the distance between the Sun and each one of the clones was within the limit of $2$ au. The shortest period was $6.3$ years. The calculated median was $\approx13,300$ years, but we registered one extreme case in which this time interval was $\approx 3.7\times 10^5$ years. The lifetime of this clone was $2.729\times10^6$ years. Therefore, it spent $13.4\%$ of its lifetime evolving in such way that the crossing of its orbit into the terrestrial region occurred. However, $31$ clones $(73.8\%)$ spent less than $1\%$ of their lifetimes in this type of orbit.

In the histogram of Fig. \ref{fig_clones_terrest}c, we present the time that each clone remained in the planetary region after leaving the terrestrial region. More than $50\%$ of the clones did not stay more than $60,000$ years. The longest time was $1.3\times10^7$ years. Except for this case, the clones that reach the terrestrial region are about to be ejected from the planetary region.

Only two clones had extreme encounters (distances $\leqslant 3\,r_{td}$) with Earth and/or Venus (Figs. \ref{fig_evolution} and \ref{fig_evolution2}). They did not have any previous encounters within $1\,r_{td}$ with the giant planets. Therefore, these are the cases in which the Typhon-Echidna system suffered close encounters with Earth and/or Venus as a binary. Figs. \ref{fig_evolution} and \ref{fig_evolution2} show the orbital evolutions and the histories of the encounters of these clones. The percentage loss of the binaries suffered by each one of these clones was obtained through the simulations performed in Section \ref{sec_bin_evol}, where it was considered a cloud of 12960 Typhon-Echidna-like systems performing each one of the registered encounters.

In one case, the binary would probably be disrupted by extreme encounters with both Earth and Venus (Fig. \ref{fig_evolution}). This body survived $5.73$ Myr and completed its evolution after a collision with the Sun. In the other case (Fig. \ref{fig_evolution2}), the binary survived extreme encounters with the giant and terrestrial planets. This Venus-crossing clone was considered ejected after $1.93$ Myr of evolution. Therefore, of the whole set of 500 clones, only one $(0.2\%)$ reached the terrestrial region as a binary and left as such. Similarly, only one $(0.2\%)$ of the clones lost its second component via close encounters with the terrestrial planets).

\section{Typhon-Echidna - past evolution}
\label{sec_past}

Backward numerical integrations of the clones of Typhon were performed in order to study the past evolution of the Typhon-Echidna system. Among the $500$ clones previously considered, we randomly selected a sample of $100$. The orbits of these clones were numerically integrated for $100$ Myr in the past. The planets and the other parameters of this simulation are the same as in Section \ref{sec_orb_evol}, except for the ejection distance. Here, a clone was considered ejected if it reached a relative distance to the Sun of $1000$ au, since it is assumed that at this distance, it is about to enter in the Oort cloud \citep{Levison2006}. The significant close encounters with any of the planets were also recorded throughout the backward integrations.

We then analysed the temporal evolutions of the semi-major axis $a$, eccentricity $e$ and inclination $i$. For each clone, we computed the locations in the $(a\times e)$ and in the $(a\times i)$ spaces
where the clones spent most of the time along the integration. The graphs presented in Fig. \ref{fig_example} exemplify how this was done. For each clone, we have the temporal orbital evolution in the $(a\times e)$ and in the $(a\times i)$ spaces, as shown in Figs. \ref{fig_example}a and \ref{fig_example}b. Those spaces were then divided assuming a regular grid, and how long the clone remained in each one of these partitions was computed. From this example, we found that the clones mostly remained within $a=42.5$ au, $e=0.265$ and $i=27.8^{\circ}$.

The results for the whole set of clones are presented in Figs. \ref{fig_past}(a) and \ref{fig_past}(b). Each point on these graphs was obtained as described above. These plots show that Typhon must have spent most of its time in the past $100$ Myr as a TNO-Centaur. From the total sample of $100$ clones, $81$ spent the past mainly as a TNO-Centaur, while $16$ spent most of their lifetime as a Centaur and only $3$ spent the majority of time as a TNO. It is possible to see a large dispersion over the whole TNO-Centaur region without any preferable locus. From the wide range of possibilities, we found that the present Typhon may have kept its current orbit for the last $100$ Myr or even may have had a Ceto-like orbit (black circles in Figs. \ref{fig_past}(a) and \ref{fig_past}(b)) in the past. On the other hand, Ceto can evolve until it becomes the current Typhon. The results show that $71\%$ of the clones were ejected towards the Oort cloud. 

Once we found that Typhon may have been a TNO-Centaur for the last hundred million years, we can expect that the close encounters of this system with the giant planets may have been as frequent in the past as we found for the forward evolution. In fact, the backward integrations show that, of the total of $100$ clones, $25$ have suffered past disruptive encounters with the giant planets (encounters within $1\,r_{td}$). In total, $10$ disruptive encounters with Neptune, $7$ with Jupiter, $4$ with Saturn and $4$ with Uranus were registered. These results show that the binary system of Typhon-Echidna could be as old as $100$ Myr, since $75\%$ of the clones would survive as binaries over this past time span.

\cite{Nol2006a} explored whether a Typhon-Echidna-like system would survive a transition from the scattered disk (TNOs with $a>50$ au and $q>30$ au.) to its current orbit. They found that there is a maximum chance of $95\%$ of this happening, considering the separation of the components of the binary $a_B=2700~km$ (this value was undetermined at the time of their study). Although their result also shows a favourable survival rate of the binary system, their values are not comparable to ours. They considered bodies that were initially placed in the scattered disk, and through forward numerical integrations, they monitored those that reached a region near the current orbit of Typhon-Echidna. In our simulations, our clones are in a very small region near the current orbit of Typhon, and the separation of Typhon-Echidna is now known, so we used much smaller values than those adopted previously.

\section{Final Comments}
\label{sec_final}
In this paper, we presented a scenario of the possible evolutions of the binary TNO-Centaur Typhon-Echidna. A system of $500$ clones of Typhon, the Sun and the planets of the Solar System (excluding Mercury) was considered. The orbits of the clones were numerically integrated for a time span of $200$ Myr. Throughout the integrations, significant close encounters of these clones with any of the planets were registered.

It was found that this system frequently crosses the orbits of the giant planets, leading to numerous 
significant and extreme close encounters with these planets. The most encountered planet was Jupiter, followed by Uranus, Saturn and Neptune. Nevertheless, we found that only $17\%$ of the clones suffered extreme encounters along the numerical integration over $<200$ Myr. The encounters suffered by those clones were then simulated considering the binary systems with a wide range of possible initial configurations. This approach allowed us to statistically analyses the probability of the disruption of the Typhon-Echidna binary system within the planetary region.  

The simulations of these encounters showed that the majority of extreme encounters between Typhon-Echidna and the giant planets were not strong enough to lead to the disruption of the binary. From the total of 619 extreme encounters registered for these planets, only approximately $22\%$ led to the disruption of the system. Thus, it was shown that it is highly probable that the binary system of Typhon-Echidna survives the close encounters with the giant planets while a Centaur.

A peculiar result that was obtained is the probability of having a binary system with components as large as the components of Typhon-Echidna that enters the terrestrial planetary region. Among the $500$ clones of the sample, $42$ reached the terrestrial region, leading to a probability of $8.4\%$ of such an event occurring during the range of the numerical integration ($<200$ Myr). The probability of Typhon-Echidna reaching the terrestrial planet region as a binary was estimated to be $3.6\%$. It is more probable that the Typhon-Echidna system would spend just a small fraction of its lifetime (less than $1\%$) in this region, and this period usually occurs at its final stage within the planetary region.

\cite{napier2015} and \cite{napieretaal} discuss that the entry of a Centaur into the terrestrial region, taking into account its fragmentation due to the sublimation or tidal disruption of the Sun or Jupiter, significantly increases the amount of Earth-crossing debris. It was estimated that a variation of two orders of magnitude in the mass of the near-Earth asteroid population is observed over a timescale of $30,000 - 100,000$ years. A major consequence of such an event is that the increase in the number of debris objects also increases the risk of a catastrophic event caused by a collision of those debris with our planet.  We highlight that the presence of a binary Centaur in the terrestrial region, as described in the present paper, may potentiate this effect. 

We also report the curious case of a binary TNO-Centaur being disrupted by close encounters with terrestrial planets. We found eight extreme encounters of Typhon-Echidna with Earth and Venus, and in two cases, those encounters were strong enough to disrupt the binary system. It is interesting to note that the presence of water ice on Typhon has been confirmed \citep{candal2010}. Therefore, the entry of Typhon-Echidna into the inner Solar System increases the possibility of this binary system presenting cometary features.

The past evolution of the Typhon-Echidna system was investigated, and it was found that Typhon must have spent most of its past as a TNO-Centaur. It was possible to see a large dispersion in the whole TNO-Centaur region without any preferable locus. By looking for the disruptive encounters in the past of the binary system, especially encounters with the giant planets, it was found that Typhon-Echidna is more likely to survive those encounters, and thus, this binary system could be as old as $100$ Myr.

Therefore, here, we presented Typhon-Echidna as an unprecedented case of a binary system comprising large cometary bodies originating from the outer Solar System that might enter the terrestrial planetary region while preserving its binarity throughout the journey.

\section{Acknowledgements}
MAG would like to thank the FWF: Project J-3588 "NEOS, impacts, origins and stable orbits''.
This work was also funded by CNPq Procs. 305737/2015-5 and 312813/2013-9 and by FAPESP Proc. 2016/24561-0. 
This support is gratefully acknowledged.

\renewcommand{\refname}{REFERENCES}

\label{lastpage}

\end{document}